\documentclass[10pt,a4paper]{article}
\usepackage{epsf}

\begin{document}

\begin{center}
{\bfseries THE PRELIMINARY RESULT FROM ($K^0_s\pi^{\pm}$) SPECTRA IN
p+A REACTION AT 10 GEV/C.}

\vskip 5mm P.Zh. Aslanyan$^{1,2,\dag}$, V.N.Emelyanenko$^1$

\vskip 5mm

{\small (1) {\it Joint Institute for Nuclear Research First
Organization }
\\
(2) {\it Yerevan State University}
\\
$\dag$ {\it E-mail: paslanian@jinr.ru}}
\end{center}

\vskip 5mm
\begin{center}
\begin{minipage}{150mm}
\centerline{\bf Abstract}

 The experimental data from 2m propane bubble chamber have been
 analyzed to search  of scalar meson $\kappa$(800) in  $K^0_s\pi^{\pm}$ spectra
  for the reaction p+A at 10 GeV/c. The $K^0_s\pi^{\pm}$ invariant mass
  spectra has shown similarly  resonant structures with
  $M_{K^0_s\pi}$=720,780 and 890 MeV/$c^2$. The M(890) peak is
  identified as the well known resonance $K^*$ from PDG.
\end{minipage}
\end{center}

\vskip 10mm

%%%%%%%%%%%%%%%%%%%%%%%%%%%%%%%%%%%%%%%%%%%%
%% MAINMATTER
%%%%%%%%%%%%%%%%%%%%%%%%%%%%%%%%%%%%%%%%%%%%

\section{Introduction}

The scalar mesons have vacuum quantum numbers and are crucial for a
full understanding of the symmetry breaking mechanisms in QCD, and
presumably also for confinement\cite{kappa}.Suggestions that the
$\sigma(600)$ and $\kappa(800)$ could be glueballs have been made.
There are theoretical arguments for why a light and broad
$\kappa(800)$($q\overline{q}$ or 4-quark state) pole can exist near
the $\kappa$ threshold and many phenomenological papers(Naive quark
model\cite{kappa}) support its existence. The $\sigma(600)$ and
$\kappa(800)$ indeed belong to the same family as the $f_0(980)$ and
$a_0(980)$ mesons (say if the $\sigma(600)$ were composed of 2 or 4
u and d type quarks) then no such mechanism would suppress the decay
$\sigma(600)\to \pi^+\pi^-$ or $\kappa(800)\to K\pi$.

 At least 3 states with quantum numbers of
$\sigma_0\to\pi^+\pi^- $ in the reaction np$\to\pi^+\pi^-$X at
$P_n$=5.2 GeV/c on 1m LHE JINR hydrogen  bubble chamber data
\cite{troy} have been found at mass ranges of 418, 511 and 757
MeV/$c^2$.

A study \cite{k892} vector mesons $K^{*\pm}$(892) in pp interactions
at 12 and 24 GeV/c by using data(280000 - events) from proton
exposure of CERN 2m hydrogen bubble chamber. Total inclusive cross
sections in pp interactions are equal to 0.27$\pm$ 0.03 and 0.04$\pm
^{0.02}_{0.03}$ for $K^{*+}$ and $K^{*-}$, respectively.
$K^0_s\pi^+$ and $K^0_s\pi^-$ invariant mass spectra has shown in
this report\cite{k892} there are small enhancement over mass regions
of 710, 800 MeV/$c^2$ and significant statistical enhancement over
mass region of $K^{*+}$(892)($K^{*\pm}$(892)) at 12 GeV/c(24 GeV/c).

\section{Experiment}

The full experimental information of more than 700000 stereo
photographs or $10^6$ p+propane inelastic interactions are used to
select of the events with $V^0$ strange particles.The invariant
masses of the observed 8657-events with $\Lambda$ hyperon
4122-events with $K_s^0$ meson  are consistent with their PDG
values\cite{v0}. The experimental total cross sections are equal to
13.3 and 4.6 mb for $\Lambda$ and $K_s^0$ production in the p+C
collisions at 10 GeV/c. From published article \cite{lk}  one can
see that the experiment is satisfactorily described by the FRITIOF
model\cite{fri1}.

\section{$K^0_s \pi^{\pm}$ spectra analysis}

For the fit of the resonance signals, the mass spectra were taken to
have the form\cite{bg} $d\sigma(M)$/dm = BG(M)+BW(M)*PS(M), where
BG,BW and PS represent background, Breit-Wigner(BW) function and
phase space, respectively. The  background has been  obtained by
three methods. The first is a polynomial (or polynomial of Legendre)
method. The second method of the randomly mixing angle between
$K^0_s$ and $\pi$ from different experimental events was described
in \cite{mix}.The third type of background has been obtained by
FRITIOF model\cite{fri1}.

The statistical significance of resonance peaks were calculated as
NP /$\sqrt{NB}$, where NB is the number of counts in the background
under the peak and NP is the number of counts in the peak above
background.The average effective mass resolution  of $K^0_s\pi$
system  is equal to 2.5 \%.

\subsection{$K^0_s\pi^+$ - spectrum}

\begin{figure}
  \epsfysize=80mm
 \epsfxsize=70mm
 \centerline{
 \epsfbox{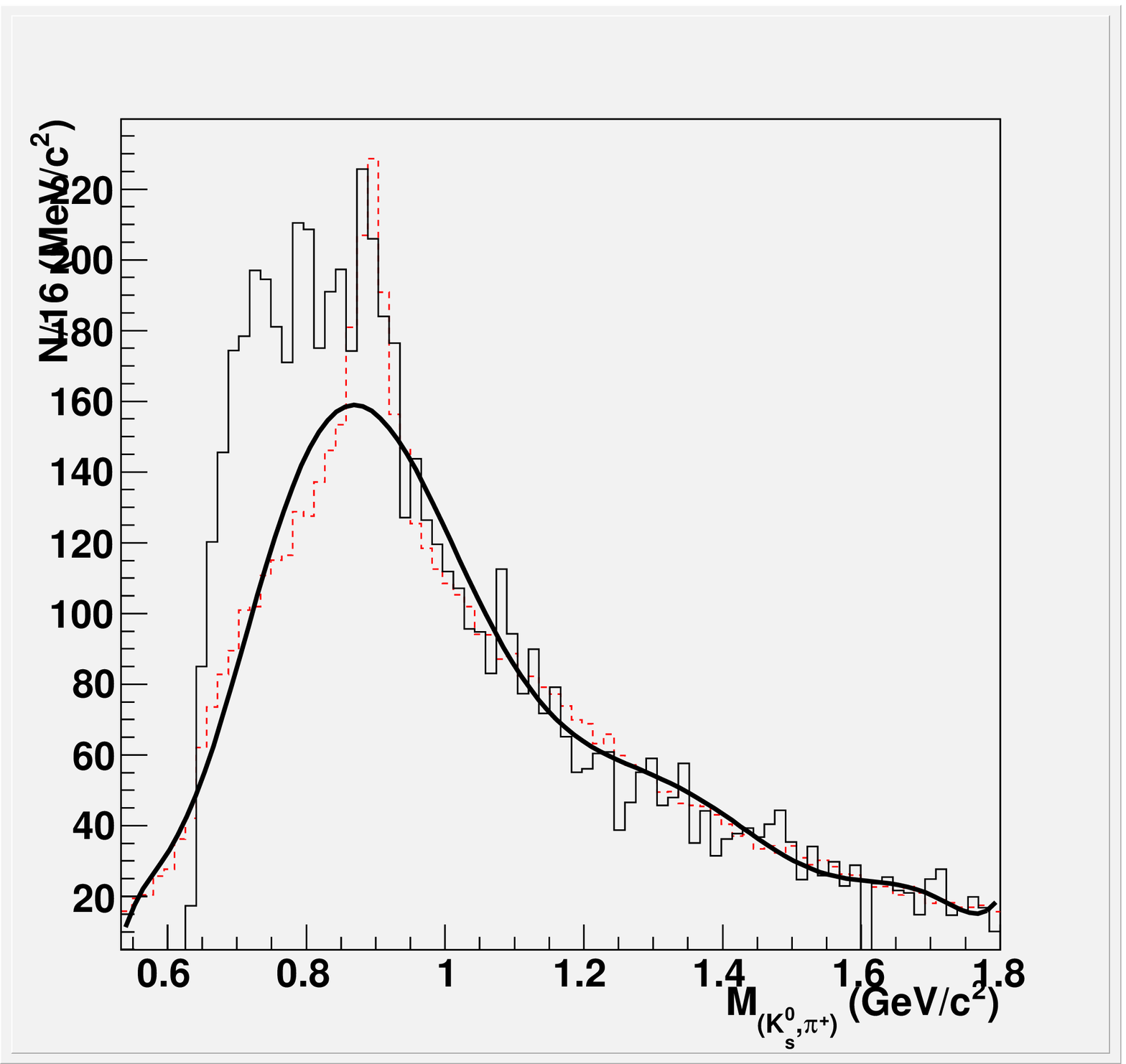}a)\epsfysize=80mm
 \epsfxsize=70mm \epsfbox{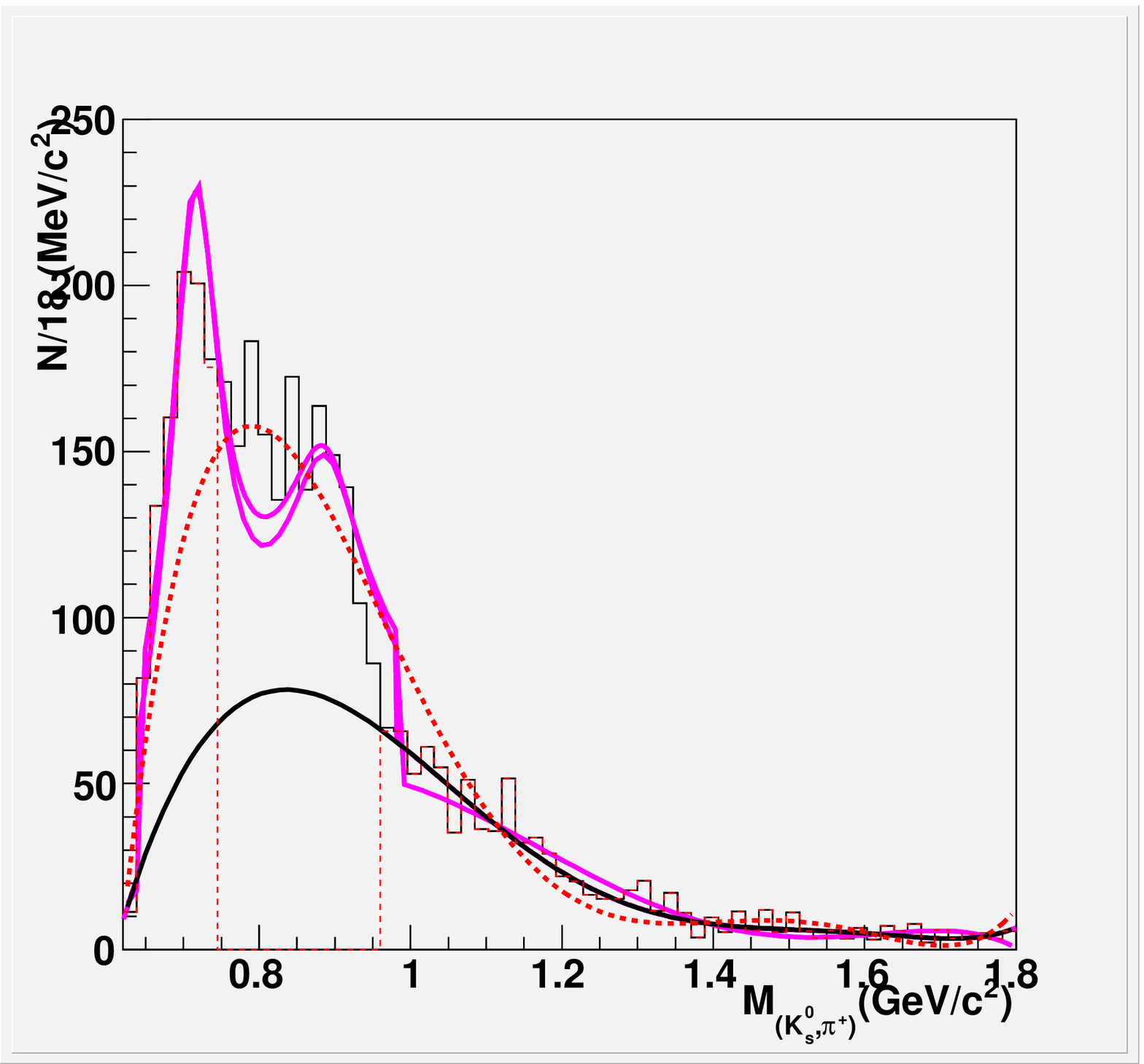}b) }
  \caption{Invariant mass distribution ( $K^0_s \pi^+$) in the
 inclusive reaction p+$C_3H_8$.  }
  \label{kpipf}
\end{figure}

Figure~\ref{kpipf}a  shows  the invariant mass distribution from all
experimental 6400($K^0_s\pi^+$ )combinations with bin sizes 16
MeV/$c^2$. In Figure~\ref{kpip10f}a the simulated dashed
distribution(94878 comb.) by FRITIOF for interaction p+propane
$\to\pi^+ K^0_s$X with experimental conditions has been normalized
to the experimental distribution. The solid curve is the backgroud
for simulated events by the FRITIOF and taken in the form of a
polynom up to the 8-th degree(Figure~\ref{kpipf}a which was agreed
with background by polynomial method. There are enhancements in mass
regions of: 720,780,840,890 and 1070 MeV/$c^{2}$. The peak M(890)in
$K^0_s\pi^+$ invariant mass spectrum  is identified as well known
resonances from PDG .

The effective mass distributions of  3259($K^0_s\pi^+$ )combinations
over the momentum range of $0.05<p_{\pi^+}<0.900$ GeV/c with bin
sizes  18 MeV/$c^2 $ is shown in Figure~\ref{kpipf}b. The
backgrounds of the experimental data were based on FRITIOF and the
polynomial method. These behaviors of backgrounds has obtained
similarly form. There are enhancements in mass regions of: 720,778
and 890 MeV/$c^{2}$. The solid curve in Figure~\ref{kpipf}b is the
sum of 2BW and background (below black solid curve) taken in the
form of a polynomial up to the 6-th degree. The dashed curve(red) is
the background by polynomial without range of 0.75< $M_{K^0_s\pi}$
<0.98 MeV/$c^2$ when a 1BW function was used.

\subsection{$K^0_s\pi^-$ - spectrum}

\begin{figure}
  \epsfysize=80mm
 \epsfxsize=70mm
 \centerline{
 \epsfbox{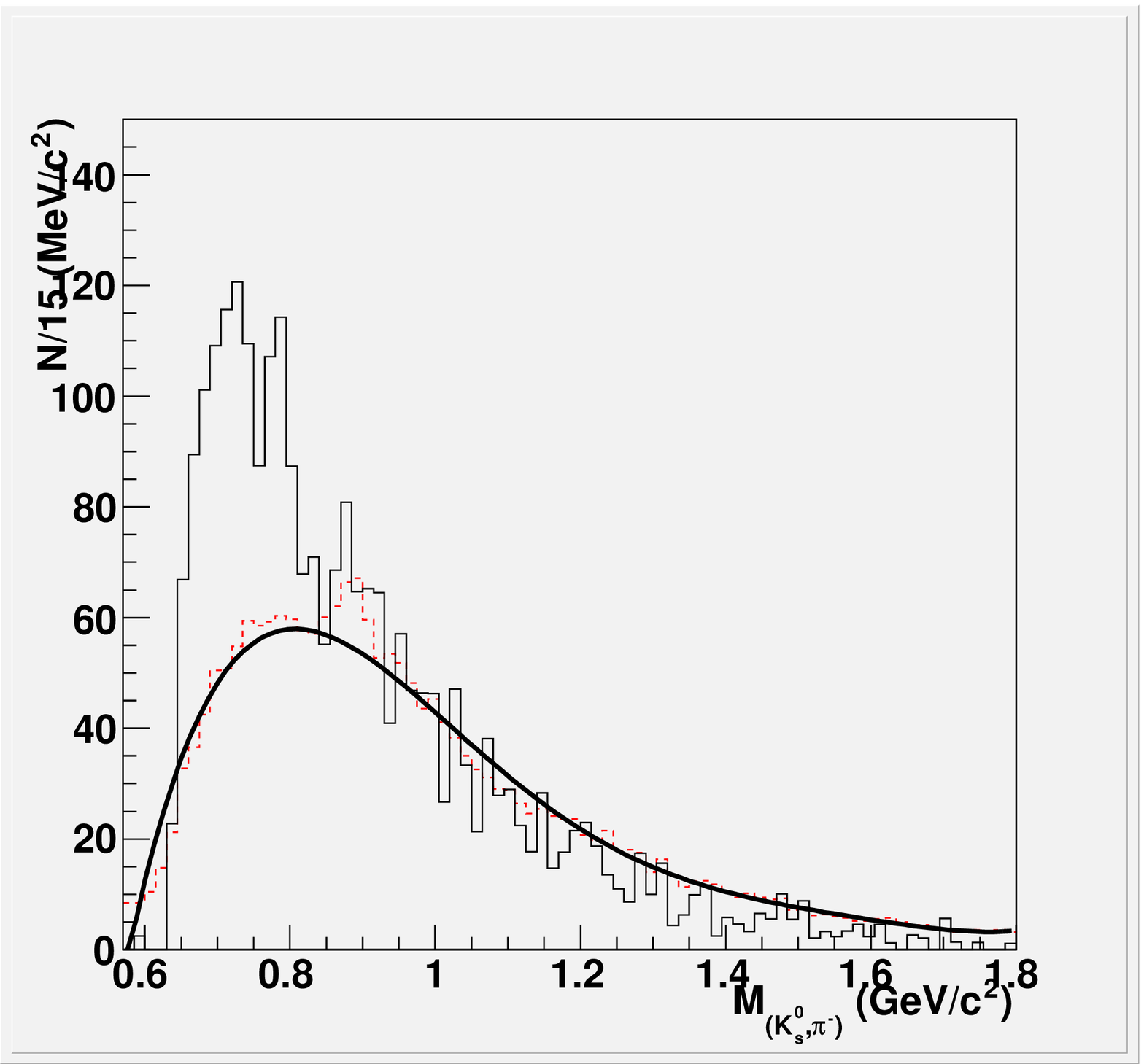}a)\epsfysize=80mm
 \epsfxsize=70mm \epsfbox{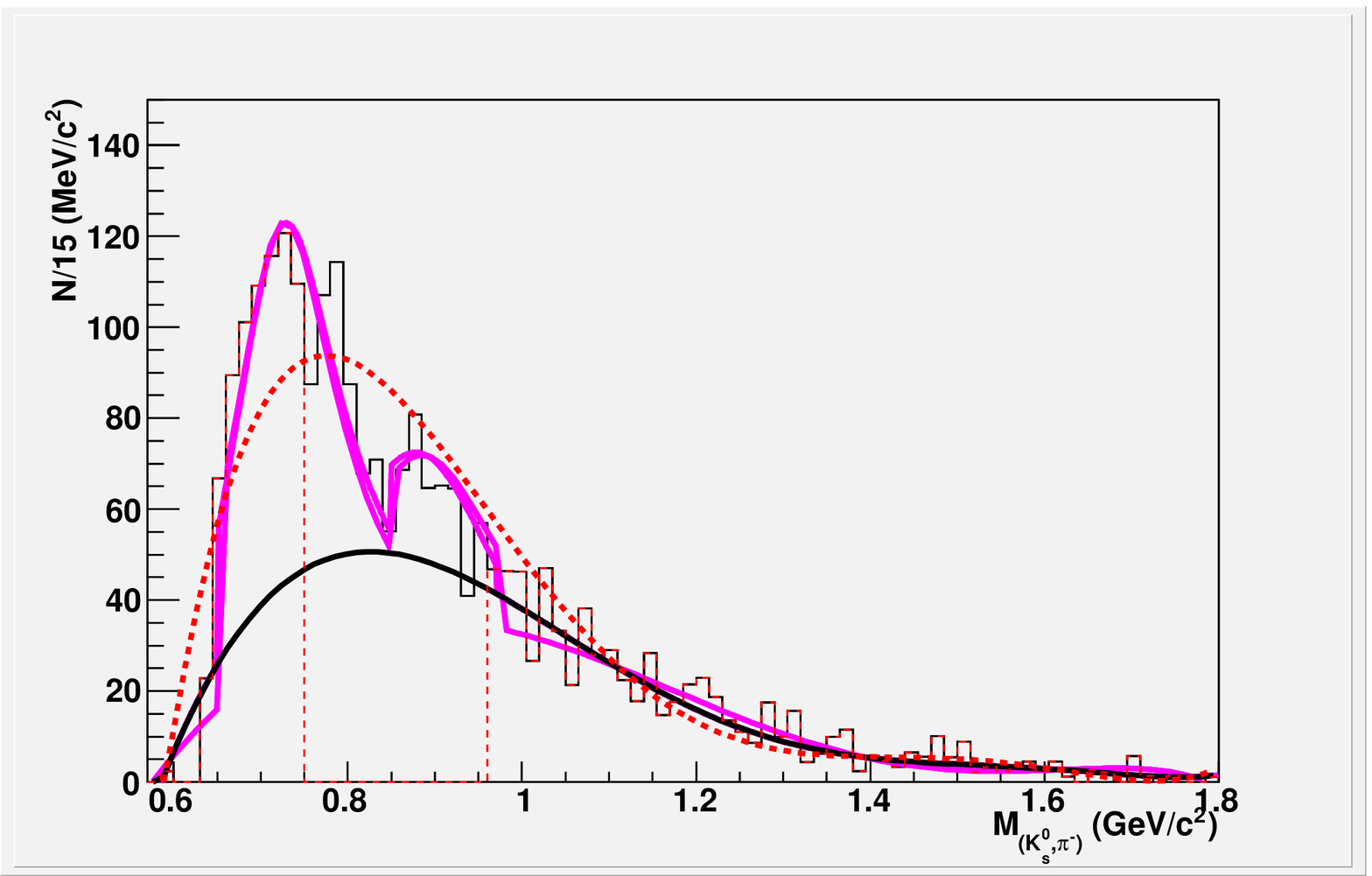}b) }
  \caption{Invariant mass distribution ( $K^0_s \pi^-$) in the
 inclusive reaction p+$C_3H_8$.}
  \label{kpim}
\end{figure}

Figure~\ref{kpim}a  shows  the invariant mass distribution
 of 2670 ($K^0_s\pi^-$ )combinations with bin size 15 MeV/$c^2$.
In Figure~\ref{kpim} the simulated background(dashed histogram) by
FRITIOF has been normalized to the experimental distribution. The
solid curve in Figure~\ref{kpim}b is the sum of 2BW and background
(below black solid curve) taken in the form of a polynomial up to
the 6-th degree. The dashed curve(red) is the background by
polynomial without range of 0.75< $M_{K^0_s\pi}$ <0.96 MeV/$c^2$
when a 1BW function was used. There are significant enhancements in
mass regions of 720,780 and 890 MeV/$c^2$(Figure~\ref{kpim}b).The
peak 890 MeV/c$^2$ in invariant mass spectrum  is identified as well
known resonances from PDG\cite{pdg}.

\section{Conclusion}

The number of  same peculiarities  were observed in the invariant
mass spectra of ($K^0_S\pi^+$) and ($K^0_S \pi^-$) subsystems  by
using PBC data for pA $\to(K^0_S\pi^{\pm})$X reactions at momentum
of 10 GeV/c in ranges of:(710-730), (770-790), (820-850), (885-895)
and (1050-1080)MeV/$c^2$. The invariant mass of  $K^0_S\pi^{\pm}$
spectra has significant enhancement in range of 890 Mev/$c^2$
($K^*(892)$ from PDG). The preliminary  interpretation of the peak
in mass ranges of 1060 Mev/$c^2$ is  sum of  reflections from well
known resonances in mass ranges of (1410-1430  Mev/$c^2$) by channel
of $K^*(892)\to (K^0_s \pi)\pi$. The preliminary total cross section
for M(730) production in p+propane interactions is estimated to be
$> 30\mu$b. The preliminary result of observation (in Table
\ref{res}) are presented  for $K^0_s\pi^{\pm}$ spectra with mass of
720 S.D. (>4.1 S.D) , 780 MeV/c2 (>2.5 S.D.) and width $\Gamma>29
MeV/c^2$ , $\approx$12 MeV/$c^2$ respectively.

%%%%%%%%%%%%%%%%%%%%%%%%%%%%%%%%%%%%%%%%%%%%
%% Sample figure:
%%
%% The option [height=...] scales the picture to the given height,
%% without it it would be printed at its nominal size
%%%%%%%%%%%%%%%%%%%%%%%%%%%%%%%%%%%%%%%%%%%%

%%%%%%%%%%%%%%%%%%%%%%%%%%%%%%%%%%%%%%%%%%%%
%% SAMPLE TABLE
%%
%% Shows the use of \tablehead and \tablenote
%% macros
%%%%%%%%%%%%%%%%%%%%%%%%%%%%%%%%%%%%%%%%%%%%

\begin{table}
\begin{tabular}{lrrrr}  \hline
Resonance & $M_{\Lambda K^0_s}$&Experimental&&The statistical  \\
Decay & MeV/$c^2$&width $\Gamma_e$&$\Gamma$&significance \\
Mode & &MeV/$c^2$&&$S.D._{max}-S.D._{min}$\\ \hline
$K^0_s\pi^{\pm}$&890&75&50&6.0-8.2\\
  $K^0_s\pi^{\pm}$&780&33&10&2.5-4.2\\
$K^0_s\pi^{\pm}$&720&>45&>27&>4.1\\

 \hline
 \end{tabular}
 \caption{The effective mass, the width($\Gamma$) and the statistical
significance of resonances produced in collisions of protons with
propane at 10 GeV/c}
 \label{res}
\end{table}

%%%%%%%%%%%%%%%%%%%%%%%%%%%%%%%%%%%%%%%%%%%
%% The following lines show an example how to produce a bibliography
%% without the help of the BibTeX program. This could be used instead
%% of the above.
%%%%%%%%%%%%%%%%%%%%%%%%%%%%%%%%%%%%%%%%%%%


\begin{thebibliography}{9}

\bibitem{kappa}C. Amsler, N.A. Tornqvist, Physics Reports 389 (2004),61–117.
\bibitem{troy}Yu. A. Troyan et al., Particles and Nuclei, Letters"  [114],2002
\bibitem{k892}K.Bochmann et al.,Nucl.Phys. B166,p.284,1980.
\bibitem{v0}P.Zh. Aslanyan and et al.,Physics of Particles and Nuclei Letters,
  Vol. 4, No. 1, 2007. P.Zh.Aslanyan et al.,JINR Commun.,E1-2005-150, 2005.
\bibitem{lk}P.Zh. Aslanyan and et al.,Physics of Particles and Nuclei Letters,
  Vol. 3, No. 5,pp. 331–334, 2006. P.Zh.Aslanyan et al.,JINR Commun.,E1-2005-149, 2005.
\bibitem{mix} V.L.Lyuboshits at al., JINR Rapid Comm., N6(74),p209, 1995.
\bibitem{fri1} FRITIOF, H. Pi, Comput. Phys.Commun. 71,173, 1992.
  A.S.Galoian et al., JINR Commun., P1-2002-54, 2002.
\bibitem{bg}M.Deutshman et al.,Nucl.Phys. B103,p.426,1970.
\end{thebibliography}
\end{document}